\documentclass[twocolumn,english,reprint]{revtex4}
\usepackage[T1]{fontenc}
\usepackage[latin9]{inputenc}
\usepackage{array}
\usepackage{textcomp}
\usepackage{amstext}
\usepackage{graphicx}

\makeatletter

\newcommand{\lyxmathsym}[1]{\ifmmode\begingroup\def\b@ld{bold}
  \text{\ifx\math@version\b@ld\bfseries\fi#1}\endgroup\else#1\fi}

\providecommand{\tabularnewline}{\\}

\@ifundefined{textcolor}{}
{%
 \definecolor{BLACK}{gray}{0}
 \definecolor{WHITE}{gray}{1}
 \definecolor{RED}{rgb}{1,0,0}
 \definecolor{GREEN}{rgb}{0,1,0}
 \definecolor{BLUE}{rgb}{0,0,1}
 \definecolor{CYAN}{cmyk}{1,0,0,0}
 \definecolor{MAGENTA}{cmyk}{0,1,0,0}
 \definecolor{YELLOW}{cmyk}{0,0,1,0}
}

\usepackage{babel}

\usepackage{babel}

\usepackage{babel}

\usepackage{babel}

\usepackage{babel}

\makeatother

\usepackage{babel}
\begin{document}

\title{Magnetization dynamics in Co$_{2}$MnGe/Al$_{2}$O$_{3}$/Co tunnel
junctions grown on different substrates}

\author{M. Belmeguenai, H. Tuzcuoglu, F. Zighem, S-M. Chérif, Y. Roussigné,
P. Moch}

\affiliation{Laboratoire des Sciences des Procédés et des Matériaux, CNRS-Université Paris XIII, 93430 Villetaneuse, France}

\author{K. Westerholt}

\affiliation{Institut für Experimentelle Physik, Ruhr-Universität Bochum, 44780
Bochum, Germany}

\author{A. El Bahoui, C. Genevois and A. Fnidiki}

\affiliation{Groupe de Physique des Matériaux, CNRS, 76801 Saint Etienne du Rouvray,
France}
\begin{abstract}
We study static and dynamic magnetic properties of Co$_{2}$MnGe (13
nm)/Al$_{2}$O$_{3}$ (3 nm)/Co (13 nm) tunnel magnetic junctions
(TMJ), deposited on various single crystalline substrates (a-plane
sapphire, MgO(100), Si(111)). The results are compared to the magnetic
properties of Co and of Co$_{2}$MnGe single films lying on sapphire
substrates. X-rays diffraction always shows a (110) orientation of
the Co$_{2}$MnGe films. Structural observations obtained by high
resolution transmission electron microscopy confirmed the high quality
of the TMJ grown on sapphire. Our vibrating sample magnetometry measurements
reveal in-plane anisotropy only in samples grown on a sapphire substrate.
Depending on the substrate, the ferromagnetic resonance spectra of
the TMJs, studied by the microstrip technique, show one or two pseudo-uniform
modes. In the case of MgO and of Si substrates only one mode is observed:
it is described by magnetic parameters (g-factor, effective magnetization,
in-plane magnetic anisotropy) derived in the frame of a simple expression
of the magnetic energy density; these parameters are practically identical
to those obtained for the Co single film. With a sapphire substrate
two modes are present: one of them does not appreciably differ from
the observed mode in the Co single film while the other one is similar
to the mode appearing in the Co$_{2}$MnGe single film: their magnetic
parameters can thus be determined independently, using a classical
model for the energy density in the absence of interlayer exchange
coupling.
\end{abstract}


\maketitle

\section{Introduction}

Tunnel magnetic junctions (TMJ) are of great interest due to their
use in magnetic memory (MRAM) {[}1{]}, in low field magnetic sensors
{[}2{]} and in microwave components for spintronics {[}3{]}. The tunnel
magnetoresistance (TMR) is very sensitive to the spin polarization
of the magnetic electrodes of TMJs and therefore, a highly spin-polarized
current source is strongly desired. A promising method for this purpose
is the use of half-metallic ferromagnets, which exhibit a half metallic
behavior with a gap separating two energy bands of opposite spin directions,
thus providing for a 100$\%$ spin polarization. Therefore, due to
their half-metallicity and to their high Curie temperature, Co-based
Heusler alloys are now used as electrodes in TMJs. Fully epitaxial
MTJs consisting of a Co-based full Heusler thin film as a lower electrode,
of a MgO tunnel barrier and of a CoFe upper electrode demonstrated
TMRs of 42$\%$ at room temperature {[}4{]}. Moreover, TMRs of up
to 220$\%$ at room temperature and of 390$\%$ at 5K have been measured
in magnetic tunnel junctions (MTJs) using Co$_{2}$FeSi$_{0.5}$Al$_{0.5}$
Heusler alloy electrodes {[}5{]}. However, up to now, the demonstrated
TMR amplitudes using Heusler alloys as magnetic electrodes in MTJs
remain lower than those using normal transition metals or their alloys.
The quality of MTJs depends strongly on the interfacial roughness,
interdiffusion and oxygen content, which in turn depend on the materials
used in the stack and on the conditions of deposition and annealing.
Therefore, the control of such parameters should enhance the half-metallicity
of electrodes and thus increase TMR values. Ferromagnetic resonance
(FMR) allows for the investigation of the dynamics of the magnetization
of single layers and of multilayers and thus for the determination
of coupling parameters and of magnetic anisotropy in relation to the
interfacial characteristics. Therefore, FMR and vibrating sample magnetometry
have been used for the investigation of static and dynamic magnetic
properties Co$_{2}$MnGe (13 nm)/Al$_{2}$O$_{3}$ (3nm)/Co (13 nm)
TMJ deposited on a-plane sapphire, MgO and Si substrates.

\section{Samples}

Co$_{2}$MnGe (13 nm)/Al$_{2}$O$_{3}$ (3nm)/Co (13 nm) TMJs have
been deposited by RF magnetron sputtering in an ultrahigh vacuum on
the following substrates: a-plane sapphire, MgO (100) and Si(111).
For comparison, single films of Co$_{2}$MnGe and Co of the same thickness
(13 nm) deposited on sapphire were also grown. All the samples, except
the Co single layer, were preliminarily covered with a 4 nm thick
vanadium-seed-layer and subsequently overlayered by a 4 nm thick gold
layer protecting them against oxidation. A more detailed description
of the sample preparation procedure can be found elsewhere {[}6{]}.
The $\theta-2\theta$ X-rays pattern (using a Cobalt line source at
$\lambda=1.78897$ Å) shown on figure 1a indicates that the Co$_{2}$MnGe
single thin film is (110) oriented. Its deduced lattice constant ($a=5.818$
Å) is slightly higher than in the bulk material ($a=5.743$ Å) {[}7,
8{]}. Figure 1b shows the cross-sectional high resolution transmission
electron micrograph of the TMJ grown on a-plane sapphire substrate.
This substrate appears as a flat single crystal exempt of roughness.
The vanadium layer is 4 nm thick and grows epitaxially on the substrate.
The interface between the substrate and the V layer is well-defined
and does not present diffusion (chemical analysis). The Co$_{2}$MnGe
film starts to grow epitaxially on V and then becomes polycristalline
above few atomic layers. The 13 nm thick Co$_{2}$MnGe layer remains
homogenous over the whole thickness without presenting amorphous areas.
Its polycrystalline nature has been confirmed by figures of poles
{[}9{]}. The 3 nm thick Al$_{2}$O$_{3}$ film is completely amorphous
with well defined interfaces and without roughness. The and is 13
nm thick Co layer is polycrystalline. The Co-Al$_{2}$O$_{3}$ interface
is smooth while the Au-Al$_{2}$O$_{3}$ is very rough and wavy.

\begin{figure}
\includegraphics[bb=75bp 370bp 440bp 560bp,clip,width=8.5cm]{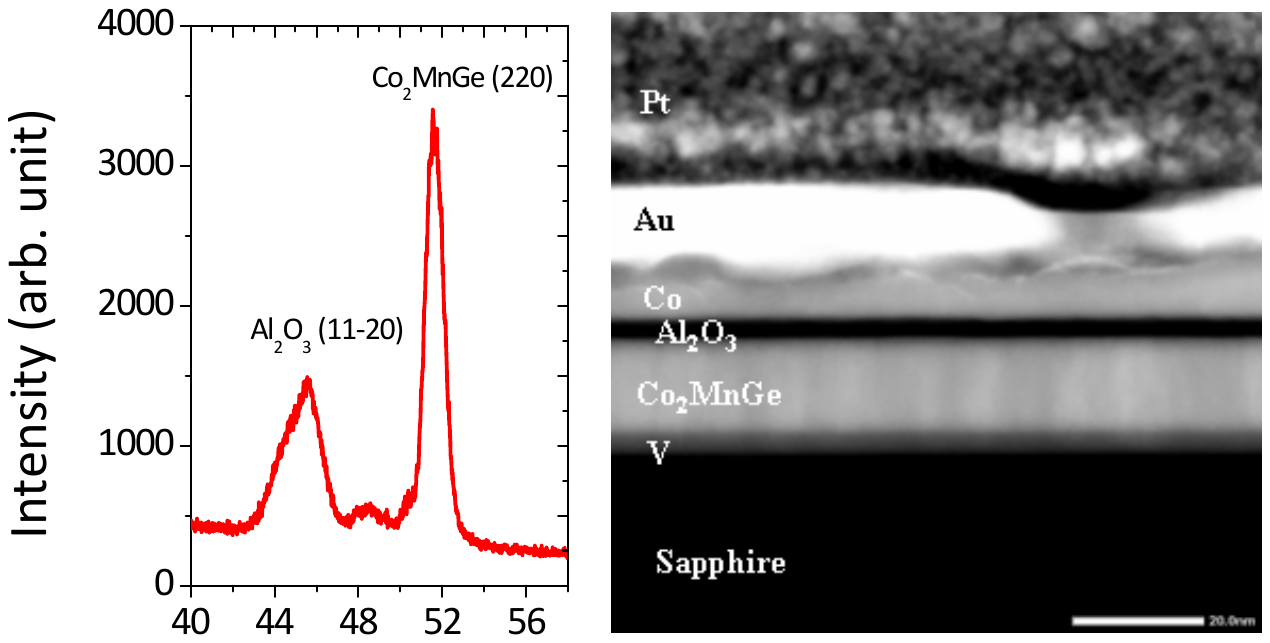}

\caption{X-rays Bragg scan using the CoK$_{\alpha}$ radiation of a 13 nm thick
Co$_{2}$MnGe thin film and Cross-sectional transmission electron
micrograph of a Co$_{2}$MnGe (13 nm)/Al$_{2}$O$_{3}$ (3nm)/Co (13
nm) TMJ grown on sapphire substrate.}
\end{figure}

\begin{figure}
\includegraphics[bb=80bp 40bp 320bp 560bp,clip,width=8.5cm]{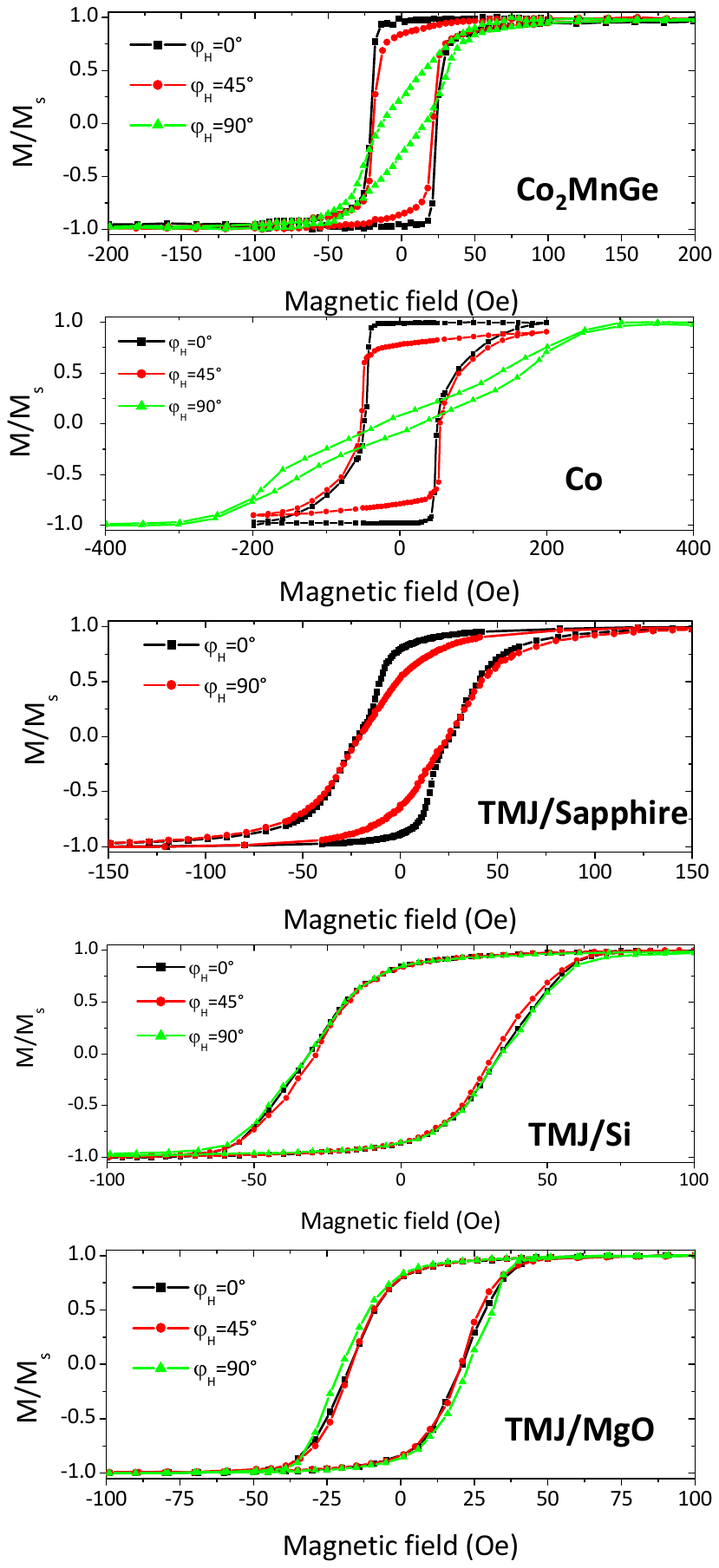}

\caption{(Color online) VSM magnetization loops of the different studied samples.
The magnetic field is applied parallel to the film surface, at various
angles ($\varphi_{H}$) with the substrate edges.}
\end{figure}

\begin{table*}
\begin{tabular}{>{\centering}p{2cm}>{\centering}p{1cm}>{\centering}p{1cm}>{\centering}p{1cm}>{\centering}p{1cm}>{\centering}p{1cm}>{\centering}p{1cm}>{\centering}p{1cm}>{\centering}p{1cm}>{\centering}p{1cm}>{\centering}p{1cm}>{\centering}p{1cm}>{\centering}p{1cm}}
\hline
Sample & \multicolumn{2}{>{\centering}p{2cm}}{$g$ } & \multicolumn{2}{>{\centering}p{2cm}}{$4\pi M{}_{\mathit{eff}}$

(kOe) } & \multicolumn{2}{>{\centering}p{1.5cm}}{$H{}_{\mathit{u}}$

(Oe)} & \multicolumn{2}{>{\centering}p{2cm}}{$H{}_{\mathit{4}}$

(Oe)} & \multicolumn{2}{>{\centering}p{2cm}}{$\varphi{}_{\mathit{u}}$

(deg.)} & \multicolumn{2}{>{\centering}p{2cm}}{$\varphi{}_{4}$

(deg.)}\tabularnewline
\hline
Co2MnGe & \multicolumn{2}{>{\centering}p{2cm}}{2.02} & \multicolumn{2}{c}{8.9 } & \multicolumn{2}{>{\centering}p{2cm}}{45} & \multicolumn{2}{>{\centering}p{2cm}}{40} & \multicolumn{2}{>{\centering}p{2cm}}{12} & \multicolumn{2}{>{\centering}p{2cm}}{0}\tabularnewline
Co & \multicolumn{2}{>{\centering}p{2cm}}{2.17} & \multicolumn{2}{>{\centering}p{2cm}}{15.5} & \multicolumn{2}{>{\centering}p{2cm}}{300 } & \multicolumn{2}{>{\centering}p{2cm}}{} & \multicolumn{2}{>{\centering}p{2cm}}{0} & \multicolumn{2}{>{\centering}p{2cm}}{}\tabularnewline
\multicolumn{1}{>{\centering}p{2cm}}{} & M1  & M2 & \multicolumn{1}{>{\centering}p{1cm}}{M1 } & M2 & \multicolumn{1}{>{\centering}p{1cm}}{M1 } & M2 & M1  & M2 & M1  & M2 & M1  & M2\tabularnewline
TMJ/saphir & 2.02 & 2.17 & 9.2 & 16 & 34 & 11 & 32 &  & -15 & 0 & 0 & \tabularnewline
TMJ/Si &  & 2.17 &  & 16 &  & 0 & \multicolumn{2}{>{\centering}p{2cm}}{} &  & 0 & \multicolumn{2}{>{\centering}p{2cm}}{}\tabularnewline
TMJ/MgO &  & 2.17 &  & 16 &  & 0 &  &  &  & 0 &  & \tabularnewline
\end{tabular}

\caption{Magnetic parameters obtained from the best fits to our FMR experimental
results. $\varphi{}_{\mathit{u}}$ and $\varphi{}_{\mathit{4}}$ are
the angles of the in-plane uniaxial and of the fourfold anisotropy
easy axes, respectively.}
\end{table*}

\begin{figure*}[t]
\includegraphics[bb=70bp 430bp 780bp 560bp,clip,width=18cm]{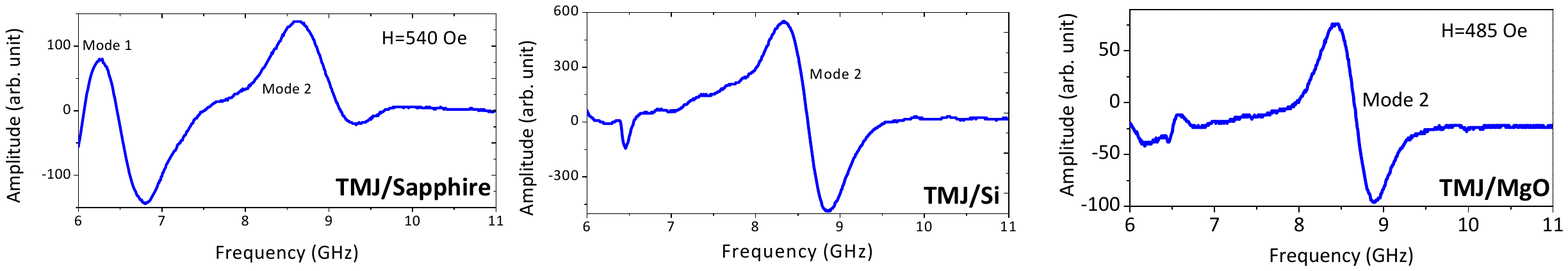}

\caption{MS-FMR spectra measured of Co$_{2}$MnGe(13nm)/Al$_{2}$O$_{3}$(3nm)/Co(13
nm) TMJs grown on a-plane sapphire, Si(111) and MgO(100) substrates.}
\end{figure*}

\section{Results and discussion}

All the reported experiments were performed at room temperature. The
static magnetic measurements were carried out using a vibrating sample
magnetometer (VSM) and the dynamic magnetic properties were investigated
with a microstrip ferromagnetic resonance MS-FMR {[}10{]} device.

\subsection{Static measurements}

In order to study the magnetic anisotropy, the hysteresis loops were
measured for all the studied samples with an in-plane magnetic field
applied along various orientations $\varphi_{H}$ (where $\varphi_{H}$
is the in-plane angle between the magnetic applied field $H$ and
one substrate edge), as shown on figure 2. The variations of the reduced
remanent magnetization ($M_{R}/M_{S}$) were then investigated as
function of $\varphi_{H}$. The magnetization loops show that single
layers and TMJ on sapphire present clearly in-plane magnetic anisotropy
and provide coercive fields ($H_{C}$) along the easy axis of 23 Oe,
48 Oe and 23 Oe in Co$_{2}$MnGe, Co and TMJ, respectively. The hysteresis
loops of TMJs on Si and MgO provide coercive fields of 33 Oe and 19
Oe, respectively, but do not depend on $\varphi_{H}$ varies, suggesting
the absence of anisotropy. The Co layer presents a high uniaxial anisotropy
of about 300 Oe, as deduced from the hard axis magnetization loop
($\varphi_{H}=90^{\circ}$), most probably induced by the interface
with the sapphire substrate {[}11{]}. The Co$_{2}$MnGe single layer
presents a complex anisotropy which cannot be simply described using
a uniaxial anisotropy and complementary experiments should be done
to clarify it. The hysteresis loop of TMJ grown on sapphire presents
a narrow plateau for small applied fields suggesting an antiparallel
alignment of the Co and Co$_{2}$MnGe magnetizations and, consequently,
only a very weak coupling of the two layers. Indeed, for a high applied
magnetic field, the two magnetizations are parallel and aligned with
it. When the sign of the applied field changes, due to the slight
difference between the coercive fields of Co and of Co$_{2}$MnGe,
the magnetizations of the two layers do not simultaneously switch,
thus leading to small plateau in the hysteresis loop. The observed
differences between the TMJs and single layers reinforce the suggestion
of an interfacial origin of the anisotropy.

\begin{figure}[h]
\includegraphics[bb=65bp 380bp 295bp 565bp,clip,width=8cm]{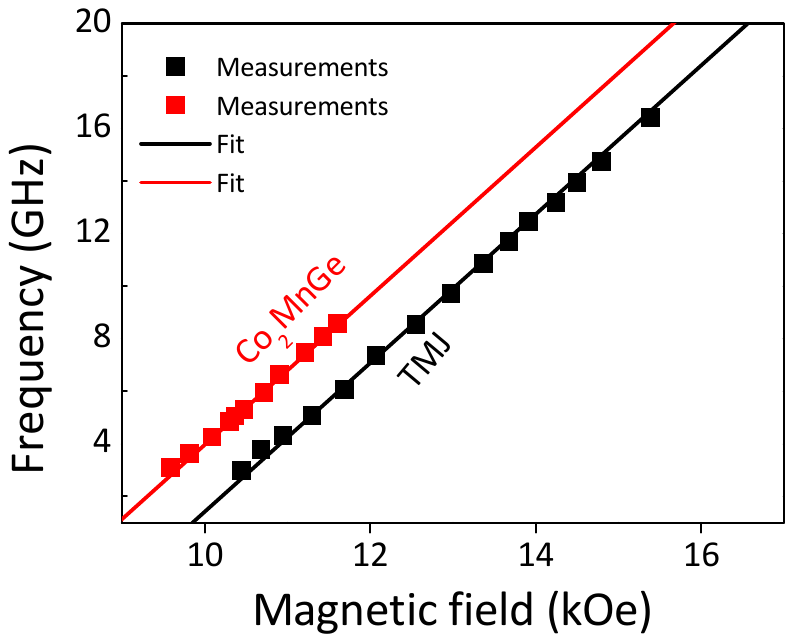}

\caption{(Color online) Field dependence of the resonance frequency of mode
1 of the 13 nm thick Co$_{2}$MnGe film and Co$_{2}$MnGe(13nm)/Al$_{2}$O$_{3}$(3nm)/Co(13
nm) TMJ grown on a-plane sapphire substrate. The magnetic field is
applied perpendicular to the film plane. The fits are obtained using
equation (3) with some parameters indicated in the text and Table
I. }
\end{figure}

\subsection{Dynamic measurements}

For each single layer, we assume a magnetic energy density which,
in addition to Zeeman, demagnetizing and exchange terms, is characterized
by the following anisotropy contribution:\\

$E=-M_{S}H(\sin\theta_{M}\sin\theta_{H}\cos(\varphi_{M}-\varphi_{H})+\cos\theta_{M}\cos\theta_{H})-(2\pi M_{S}^{2}-K_{\bot})\sin^{2}\theta_{M}-\frac{1}{2}(1+cos(2(\varphi_{M}-\varphi_{u}))K_{u}\sin^{2}\theta_{M}-\frac{1}{8}(3+\cos4(\varphi_{M}-\varphi_{4}))K_{4}\sin^{4}\theta_{M}$
(1)

For an in-plane applied magnetic field $H$, the studied model provides
the following expression for the frequencies of the experimentally
observable magnetic modes:\\

$F\text{\texttwosuperior}=\left(\frac{\gamma}{2\pi}\right)^{2}(H\cos(\varphi_{H}-\varphi_{M})+\frac{2K_{4}}{M_{S}}\cos4(\varphi_{M}-\varphi_{4})+\frac{2K_{u}}{M_{S}}\cos2(\varphi_{M}-\varphi_{u}))\times(H\cos(\varphi_{H}-\varphi_{M})+4\pi M_{eff}+\frac{K_{4}}{2M_{S}}(3+\cos4(\varphi_{M}-\varphi_{4}))+\frac{K_{u}}{M_{S}}(1+\cos2(\varphi_{M}-\varphi_{u}))$
(2)

In the above expression \textit{$(\gamma/2\pi)=g\times1.397\times10^{6}$}
s$^{-1}$.Oe$^{-1}$ is the gyromagnetic factor and $\varphi_{M}$
represents the in-plane (referring to one substrate edge) angle defining
the direction of the magnetization $M_{S}$; $\varphi_{u}$ and $\varphi_{4}$
stand for the angles of an easy uniaxial planar axis and of an easy
planar fourfold axis, respectively, with this substrate edge. $K_{u}$,
$K_{4}$and $K_{\bot}$ are the in-plane uniaxial, the fourfold and
the out-of-plane uniaxial anisotropy constant, respectively. In addition
we state that $K_{u}$ and $K_{4}$ are positive. It is often convenient
to introduce the effective magnetization $4\pi M_{eff}=4\pi M_{s}-2K_{\perp}/M_{s}$,
the uniaxial in-plane anisotropy field $H_{u}\lyxmathsym{ }=\lyxmathsym{ }2K_{u}/M_{s}$
and the fourfold in-plane anisotropy field $H_{4}\lyxmathsym{ }=\lyxmathsym{ }4K_{4}/M_{s}$
. In the case of an out-of-plane perpendicular applied magnetic field,
the resonance frequency is given by:
\[
F_{\perp}=\frac{\gamma}{2\pi}(H-4\pi M_{eff})
\]

\begin{figure}[h]
\includegraphics[bb=60bp 30bp 300bp 560bp,clip,width=7.5cm]{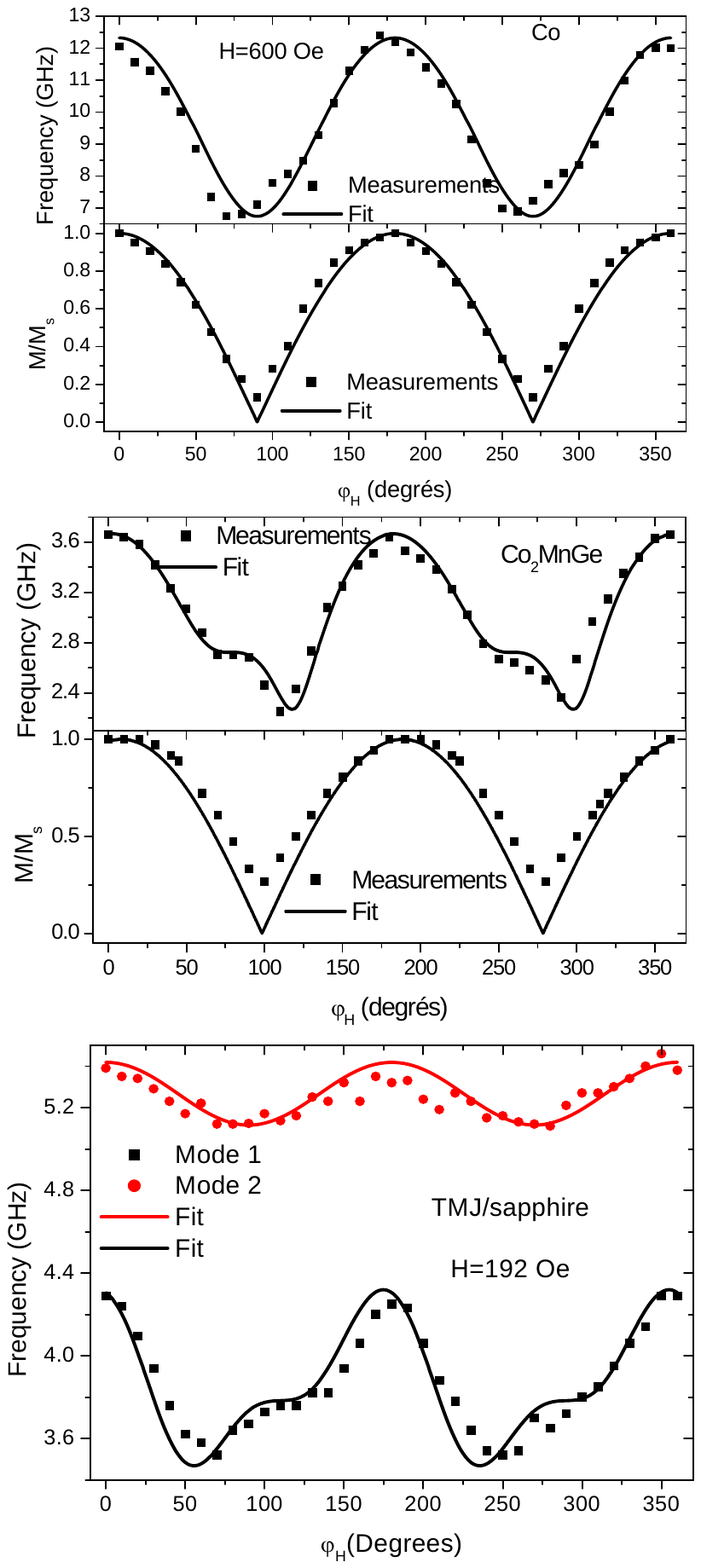}

\caption{In-plane angular dependences of the resonance frequency of modes 1
and 2 and reduced remanent magnetization of the Co, Co$_{2}$MnGe
single layers and Co$_{2}$MnGe(13nm)/Al$_{2}$O$_{3}$(3nm)/Co(13
nm) TMJ grown on a-plane sapphire substrate. The full lines are obtained
from the energy minimization (reduced magnetization) and using equation
(2) (frequency) with the parameters indicated in Table I. }
\end{figure}

\begin{figure}[h]
\includegraphics[bb=65bp 375bp 315bp 565bp,clip,width=8cm]{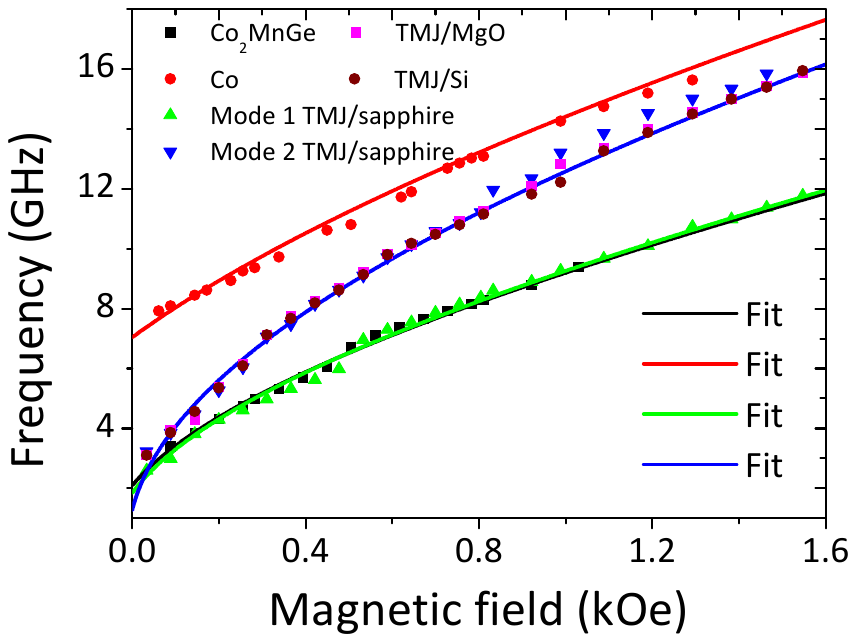}

\caption{(Color online) Field dependence of the resonance frequency of modes
1 and 2 of Co, Co$_{2}$MnGe single layers and Co$_{2}$MnGe(13nm)/Al$_{2}$O$_{3}$(3nm)/Co(13
nm) TMJ grown on a-plane sapphire, Si(111) and MgO(100) substrates.
The magnetic field is applied in the film plane. The fits are obtained
using equation (2) with parameters indicated in the text and Table
I. }
\end{figure}

MS-FMR spectra have been recorded for in-plane and perpendicular applied
fields, of variable amplitudes, for each sample. In the single films
one resonance mode is observed, as expected: it is well described
by the above model (figures 5 and 6 for Co, figures 4, 5 and 6 for
Co$_{2}$MnGe). The related fitted magnetic parameters are given in
Table I for Co and for Co$_{2}$MnGe. Notice that a uniaxial term
well describes the in-plane anisotropy in the Co sample while both
uniaxial and fourfold in-plane anisotropy are requested to give account
for the data concerning Co$_{2}$MnGe. In addition, the derived in-plane
anisotropy characteristics are consistent with the above mentioned
angular variation of the reduced remanent magnetization ($M_{R}/M_{S}$).
In the TMJs two resonance modes should be present: neglecting the
interlayers coupling the first one (mode 1) is described by magnetic
parameters near of the derived ones for the Co$_{2}$MnGe single film
and the second one (mode 2) is described by magnetic parameters near
of the derived ones for the Co single film. This expected behavior
is only observed in the TMJ grown on the sapphire substrate (figure
3). In the other TMJs the resonance study detects only a unique mode
corresponding to mode 2. The absence of mode 1 reflects their less
quality compared to TMJ on sapphire, which, due to its weak intensity,
prevents its detection. In fact, the growth conditions Co$_{2}$MnGe
layer are surely optimal for TMJ on sapphire but most probably not
for the other TMJs. In the three studied TMJs the fitted effective
g factor and the effective demagnetizing field derived from the field
dependence of the frequency of mode 2 are nearly identical to the
obtained ones in the Co single film: 2.17 and 16 kOe, to compare to
2.17 and 15.2 kOe, respectively. Concerning mode1, observed in TMJ
on sapphire, the fitted effective g factor and the effective demagnetizing
field are again nearly identical to the obtained ones in the Co$_{2}$MnGe
single film: 2.02 and 9.2 kOe, to compare to 2.02 and 8.9 kOe, respectively.
Finally, the in-plane anisotropy of mode is found to be uniaxial for
mode 2, while in mode 1 uniaxial and fourfold terms provide comparable
contributions.

\section{Conclusion}

Co$_{2}$MnGe (13 nm) / Al$_{2}$O$_{3}$ (3nm) / Co (13 nm) TMJ deposited
on a-plane sapphire, MgO and Si substrates have been prepared. For
comparison, single films of Co$_{2}$MnGe and Co of the same 13 nm
thickness been also deposited on sapphire. Their structural, static
and dynamic magnetic properties have been studied. Our X-rays diffraction
measurements show that the Co$_{2}$MnGe film is (110) oriented. The
good quality of the TMJ on sapphire has been checked via high resolution
transmission electron microscopy. The VSM measurements showed composited
anisotropy in the case of TMJ on sapphire. However, no anisotropy
has been revealed by the VSM hysteresis loops in the case of the other
TMJs. A high uniaxial anisotropyis observed in the Co film, most probably
induced the interface with sapphire substrate. MS-FMR measurements
revealed two modes for TMJ on sapphire, attributed to the Co$_{2}$MnGe
and to the Co layer and one mode for the other TMJs, attributed and
Co layer. The magnetic properties are strongly dependent on the interfaces
quality and growth condition and can be used to check the TMJ quality.
A good fit of all the experimental data, using appropriate values
of the magnetic coefficients describing the free energy, is obtained.

\end{document}